\documentclass[%
 aip,
 jmp,%
 amsmath,amssymb,
 reprint,%
]{revtex4-1}

\usepackage{graphicx}% Include figure files
\usepackage{dcolumn}% Align table columns on decimal point
\usepackage{bm}% bold math
\begin{document}
\title{Electrical modulation of the edge channel transport in topological insulators coupled to
ferromagnetic leads}

\author{Yuan Li}
\email{liyuan@hdu.edu.cn} \affiliation{Department of Electronic and
Computer Engineering, Information Storage Materials Laboratory,
National University of Singapore, 1 Engineering Drive 3, Singapore
117576, Singapore} \affiliation{Department of Physics, Hangzhou
Dianzi University, Hangzhou 310018, P. R. China}
\author{ M. B. A. Jalil}
\affiliation{Department of Electronic and Computer Engineering,
Information Storage Materials Laboratory, National University of
Singapore, 1 Engineering Drive 3, Singapore 117576, Singapore}
\affiliation{ Data Storage Institute, DSI Building, 5 Engineering
Drive 1 (Off Kent Ridge Crescent), National University of Singapore,
Singapore 117608, Singapore}
\author{Seng Ghee Tan}
\affiliation{ Data Storage Institute, DSI Building, 5 Engineering
Drive 1 (Off Kent Ridge Crescent), National University of Singapore,
Singapore 117608, Singapore}
\author{GuangHui Zhou}
\affiliation{ Department of Physics and Key
Laboratory for Low-Dimensional Quantum Structures and Manipulation
(Ministry of Education), Hunan Normal University, Changsha 410081,
China}

\begin{abstract}
The counterpropagating edge states of a two-dimensional topological insulator (TI) carry electrons
of opposite spins. We investigate the transport properties of edge states in a two-dimensional
TI which is contacted to ferromagnetic leads. The application of a side-gate voltage induces a
constriction or quantum point contact (QPC) which couples the two edge channels. The transport
properties of the system is calculated via the Keldysh nonequilibrium Green's function method.
We found that inter-edge spin-flip coupling can significantly enhance (suppress) the charge current
when the magnetization of the leads are anti-parallel (parallel) to one another. On the other hand,
spin-conserving inter-edge coupling generally reduces the current by backscattering regardless of
the magnetization configuration. The charge current and the conductance as a function of the bias
voltage, also exhibit similar trends with respect to spin-flip coupling strength, for both parallel
and anti-parallel configurations. Hence, gate voltage modulation of edge states via a QPC can provide
a means of modulating the spin or charge current flow in TI-based spintronics devices.
\end{abstract}
\date{\today}

\maketitle
\section{introduction}
Topological insulators (TIs) are electronic materials that possess
an insulating bulk gap and topologically-protected conducting states on their
edges or surfaces \cite{Kane,Bernevig}. $\mathrm{Bi_2Se_3}$ and related materials with gapless surface
states\cite{Hsieh,Zhang,Xia}, e.g., $\mathrm{Bi_2Te_3}$ and $\mathrm{Sb_2Te_3}$, have recently attracted
extensive attention in the condensed matter and device physics communities\cite{Qi,Hossain,Shahil,Fujita}.
Thin TI films were proposed for application in the magnetic memory\cite{Moore} due to their extremely high
surface-to-volume ratio\cite{Tang}. Accordingly, a mechanical exfoliation method was developed for preparation
of thin TI films with significant surface transport\cite{Teweldebrhan,Teweldebrhan2,Goyal}.

On the other hand, the two-dimensional version of
topological insulators is also known as a quantum spin hall
insulator, consisting of edge states that behave as perfect one-dimensional
counterpropagating (helical) channels. These channels are well-localized along the edges,
carry electrons with opposite spins, and
are protected from nonmagnetic impurity scattering by time-reversal symmetry.
The peculiar properties of the helical edge states have been
demonstrated theoretically and experimentally in
$\mathrm{HgTe/CdTe}$ quantum
wells \cite{Bernevig2,Konig2,Roth,Buttiker}, and have been predicted to
exist in various other materials \cite{Fu,Murakami,Liu,Knez,Shitade,Wada}. Due to the robust spin property of these
edge channels, two-dimensional topological insulators are promising candidates for
spintronic devices.

In any future spintronic device, it is desirable to achieve electrical (rather than magnetic field)
manipulation of spin transport. However, the robustness of the helical edge-state
transport to nonmagnetic impurity scattering and local perturbation, means that it is
not easy to electrically modulate the edge-state transport in TI-based devices.
Nonetheless, investigations into electrical control of edge channel transport in TIs have
gathered pace recently. For instance, due to the finite size effect \cite{Zhou}, the inter-edge tunneling arising
from the overlap between states from opposite edges can open up an energy gap, and modify the conductance of the
system. The amplitude of the inter-edge tunneling is determined by the finite decay length of the helical edge
states into the bulk \cite{Hou,Strom,Tanaka,Teo,Zyuzin}. This phenomenon may provide a means to adjust the
transport property of edge states, via finite size effect of
the quantum point contact \cite{Zhang2}, interferometry of the edge
states \cite{Dolcini}, and nanoscale engineering of the edge geometry, e.g., by patterning a constriction
in a HgTe heterostructure~\cite{Krueckl}.

In this article we propose the design of a TI-based spintronics
device which incorporates a two-dimensional topological insulator (2D-TI),
e.g. a $\mathrm{HgTe/CdTe}$ quantum well, contacted to two ferromagnetic leads.
We shall focus on the electrical manipulation of the transport properties by
inducing inter-edge tunneling through an electrically induced quantum
point contact. This can be achieved by applying external voltage via
a pair of split gate along the transverse direction of the device. We apply
the nonequilibrium Keldysh Green's function method to calculate the
charge current flowing through the TI-based device. It is found that
conductance of such device can be significantly enhanced by the
presence of inter-edge spin-flip tunneling when the leads are highly spin
asymmetric and their magnetizations are in the antiparallel
configuration. Thus, the proposed setup can potentially realize
an electrically controlled TI device in which the conductance varies
according to the inter-edge tunneling coupling and the magnetization configurations
of the leads.

The organization of the rest of the paper is as follows. In Sec.
\ref{model}, the  Hamiltonian of the system is introduced, and the
current formula is derived based on the nonequilibrium Green's
function method. In Sec. \ref{numerical}, we present the results of
our numerical calculation of the charge current and the differential
conductance in the presence of inter-edge tunneling couplings
modulated by the split-gate voltage. Finally, a brief summary is
given in Sec. \ref{summary}.
\section{model and formula}\label{model}
\begin{figure}[t]
\begin{center}
\includegraphics[width=6.8cm]{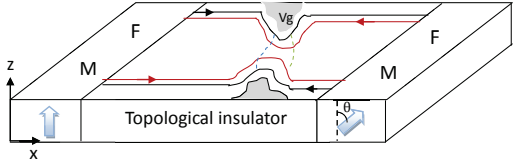}
\caption{\label{fig:device}  Schematic description of the proposed
two-terminal setup, where edge states flow at the top and bottom
boundaries of the topological insulator. There exists a QPC in the
central region which can be controlled by the split-gate voltage
$V_g$. Black (red) lines refer to spin-up (spin-down) edge channels,
which is assumed to be polarized along the $z$ direction. Application
of the split gate voltage $V_g$ reduces the spatial separation of the
two edge states, thus increasing the inter-edge tunnel coupling at the
QPC region, which is depicted by the blue (green) dashed lines.}
\end{center}
\end{figure}
We consider a TI-based device consisting of a channel made of a 2D-TI bar
sandwiched between two ferromagnetic leads [see Fig.~\ref{fig:device}].
We assume the presence of a gate-bias induced QPC, across which inter-edge
coupling occurs. In this paper, we use the low energy effective theory of the edge states to
investigate the transport property of the QSH system~\cite{Dolcini,Maciejko}.
This system can be described by the following Hamiltonian:
\begin{eqnarray}
H=H_L+H_R+H_C+H_T,
\end{eqnarray}
where $H_C$ is the effective Hamiltonian describing the QSH edge
states
\begin{eqnarray}
H_C&=&\sum_{\beta k\sigma}(\eta_\beta \eta_\sigma v k)c^\dag_{\beta
k\sigma}c_{\beta
k\sigma}+\sum_{k\sigma}[V_{c1}c^\dag_{tk\sigma}c_{bk\sigma}\nonumber\\
&&\hspace{5mm
}+\eta_\sigma
V_{c2}c^\dag_{tk\sigma}c_{bk\bar{\sigma}}+c.c.].
\end{eqnarray}
In the above, $c^\dag_{\beta k\sigma}(c_{\beta k\sigma})$ is the creation
(annihilation) operator of spin $\sigma=\uparrow$ ($\downarrow$) for
electron constrained along the $\beta$ edges, where the edge index $\beta=t,b$
refers to the top- and bottom-edge channels. The first term in $H_C$ describes the left-
and right-moving QSH edge states, while the second and third terms represent, respectively,
the spin-conserving and spin-flip couplings between the two
edge channels. Accordingly, $v$ is
the edge state velocity, $V_{c1}$ and $V_{c2}$ are, respectively, the spin-conserving and spin-flip tunnel coupling
strengths between the edge states. These can be modulated by means of the split-gate voltage $V_g$
which modifies the width of the QPC~\cite{Dolcini}. The edge and spin states are characterized by the
following indexes: $\eta_{t/b}=\pm 1$ and $\eta_{\uparrow/\downarrow}=\pm 1$. In general, in order to
preserve time-reverse symmetry, only the two types of coupling represented by $V_{c1}$ and $V_{c2}$ are allowed \cite{Bernevig}.
In the device operation, an electron is injected from the source
electrode and flows through along one of the edges of the topological insulator.
An important point to note is that under nonequilibrium condition, the total current is contributed by electrons with different
wavevector $\boldsymbol{k}$ corresponding to the conduction ``window" between the electrochemical potentials of the
source and drain electrodes. However, for simplicity, one can assume the
coupling constants to be independent of $\boldsymbol{k}$ and energy over the conduction window (this is especially valid for small source-
drain bias).

The Hamiltonian $H_L$ and $H_R$, which describe the left and right
ferromagnetic leads, and $H_T$ which describes the coupling between the FM leads
and the central TI region are given by:
\begin{eqnarray}
H_{\alpha=L,R}&=&\sum_{k\sigma}(\epsilon_{k\alpha}+\eta_\sigma
M)a^\dag_{k\alpha\sigma}a_{k\alpha\sigma}\nonumber\\
&=&\sum_{k\sigma}\epsilon_{k\alpha\sigma}a^\dag_{k\alpha\sigma}a_{k\alpha\sigma},
\end{eqnarray}
and
\begin{eqnarray}
H_T&=&\sum_{k\beta\sigma}\Big[V_{kL,
\beta\sigma}a^\dag_{kL\sigma}c_{\beta k
\sigma}+V_{kR,\beta\sigma}(\eta_\sigma\cos\frac{\theta}{2}a^\dag_{kR\sigma}\nonumber\\
&&\hspace{5mm}+\sin\frac{\theta}{2}a^\dag_{kR\bar{\sigma}})c_{\beta
k\sigma}+c.c.\Big].
\end{eqnarray}
$(\epsilon_{k\alpha}+\eta_\sigma
M)$ is the energy of conduction electrons in
the $\alpha$ lead, and is characterized by the amplitude $k$ of the wave
vector, and the amplitude $M$ and orientation $\eta_\sigma$ of the
magnetization of the FM lead. $a^\dag_{k\alpha\sigma}$
and $a_{k\alpha\sigma}$ denote the creation and annihilation operators
of electrons with spin $\sigma$ in lead $\alpha$, while
$V_{k\alpha, \beta\sigma}$ denotes the tunnel coupling between the edge state of the TI
and the lead.

The charge current can be calculated using the standard Keldysh
nonequilibrium Green's function method \cite{Meir}, namely,

\begin{eqnarray}
\label{current}
J_{\alpha}=\frac{ie}{\hbar}\int
\frac{d\epsilon}{2\pi}\mathrm{Tr}\{\mathbf{\Gamma}_{\alpha}[\mathbf{G}^<(\epsilon)
+f_{\alpha}(\epsilon)(\mathbf{G}^r(\epsilon)-\mathbf{G}^a(\epsilon))]\},\nonumber\\
\end{eqnarray}
where
$f_\alpha(\epsilon)=\{\exp[(\epsilon-\mu_\alpha)/k_BT]+1\}^{-1}$ is
the Fermi-distribution function, and the trace is over both the spin
and edge-state indexes. The linewidth $\mathbf{\Gamma}_{\alpha}(\epsilon)$ is represented by a
$(4\times 4)$ matrix in spin space and edge-state space, which can
be written as
\begin{eqnarray*} \mathbf{\Gamma}_{\alpha}(\epsilon)=\left( {\begin{array}{*{100}c}
\mathbf{\Gamma}_{\alpha}^t&\mathbf{0} \\[3mm]
\mathbf{0}&\mathbf{\Gamma}_{\alpha}^b
\end{array}} \right),\hspace{5mm}\mathbf{\Gamma}_{\alpha}^\beta=U_\alpha^\dag\left( {\begin{array}{*{100}c}
\Gamma_{\alpha\uparrow}^\beta& 0 \\[3mm]
0&\Gamma_{\alpha\downarrow}^\beta
\end{array}} \right)U_\alpha,
\end{eqnarray*}
with
\begin{eqnarray*}
U_\alpha=\left( {\begin{array}{*{100}c}
\cos(\theta_\alpha/2)&\sin(\theta_\alpha/2)\\[3mm]
-\sin(\theta_\alpha/2)&\cos(\theta_\alpha/2)
\end{array}} \right).
\end{eqnarray*}
Here, $\theta_\alpha$ denotes the magnetization direction in lead $\alpha$, so that we have $\theta_L=0$ and
$\theta_R=\theta$, i.e., fixed (variable) magnetization orientation for lead $\alpha=L$ ($R$). The components of the
linewidth matrix are related to the tunnel couplings as follows: $\Gamma_{\alpha\sigma}^{\beta}(\epsilon)=2\pi\sum_k
|V_{k\alpha,\beta\sigma}|^2 \delta(\epsilon-\varepsilon_{k\alpha
\sigma})$. In the following numerical
calculation, we assume $\Gamma_{\alpha
\uparrow/\downarrow}^{\beta}(\epsilon)=\Gamma_{\alpha
\uparrow/\downarrow}^{\beta}\theta(D-| \epsilon|)$ and
$\Gamma_{\alpha \uparrow/\downarrow}^{\beta}=\xi_\alpha\Gamma_0(1\pm
p_\alpha)$, where $D=500$ is the bandwidth, $\Gamma_0=1$ is the unit of
energy, $p_\alpha$ denotes the spin asymmetry factor for
tunneling across the barrier, and $\xi_\alpha$ is the asymmetry in
the TI-lead coupling between the left and right barriers. Assuming the junctions are
symmetrical and the leads are made of the same FM material, we have
$\xi_L=\xi_R=1$ and $p_L=p_R=p$. The retarded Green's function $\mathbf{G}^{r}(\epsilon)$ in the current formula
of Eq. \eqref{current} is also a
$(4\times4)$ matrix, which can be solved via the
standard equation of motion technique~\cite{Meir}, which yield the following Dyson's equation
\begin{eqnarray}\label{retarded}
\mathbf{G}^r(\epsilon)&=\mathbf{G}^{r,0}(\epsilon)+\mathbf{G}^{r,0}(\epsilon)(\mathbf{\Sigma}^{r(L)}
+\mathbf{\Sigma}^{r(R)}+\mathbf{V}_c)\mathbf{G}^r(\epsilon).\nonumber\\
\end{eqnarray}
$G^{r,0}(\epsilon)$ describes the edge states of the
topological insulator in the absence of TI-lead and inter-edge
tunneling couplings, i.e.,
\begin{widetext}
\begin{eqnarray}
G^{r,0}(\epsilon)=\left( {\begin{array}{*{100}c}
\displaystyle\frac{1}{(\epsilon-vk_0\eta_V+i\eta)}&0&0&0\\[3mm]
0&\displaystyle\frac{1}{(\epsilon+vk_0\eta_V+i\eta)}&0&0\\[3mm]
0&0&\displaystyle\frac{1}{(\epsilon+vk_0\eta_V+i\eta)}&0\\[3mm]
0&0&0&\displaystyle\frac{1}{(\epsilon-vk_0\eta_V+i\eta)}
\end{array}} \right),
\end{eqnarray}
\end{widetext}
where $k_0$ denotes the wavevector associated with the Fermi
energy, $V=(\mu_L-\mu_R)$ is the bias voltage and $\eta_V=\pm 1$
is adopted to describe the helical
property of the edge channels.
$\mathbf{\Sigma}^{r/a(\alpha)}=(\mp
i/2)\mathbf{\Gamma}_{\alpha}(\epsilon)$ are the self-energies due to the lead $\alpha$, while the
matrix $\mathbf{V}_c$ denotes the self-energy due to inter-edge tunneling, and is explicitly given by
\begin{eqnarray}
\mathbf{V}_c=\left( {\begin{array}{*{100}c}
0&0&V_{c1}&V_{c2}\\[3mm]
0&0&-V_{c2}&V_{c1}\\[3mm]
V_{c1}&-V_{c2}&0&0\\[3mm]
V_{c2}&V_{c1}&0&0
\end{array}} \right).
\end{eqnarray}
Finally, the lesser Green's function $\mathbf{G}^<$ in Eq. \eqref{current} can be calculated using the Keldysh
equation:
$\mathbf{G}^<(\epsilon)=\mathbf{G}^r(\epsilon)(\mathbf{\Sigma}^{<(L)}+\mathbf{\Sigma}^{<(R)})\mathbf{G}^a(\epsilon)$,
with the scattering function being given by
$\mathbf{\Sigma}^{<(\alpha)}=if_\alpha\mathbf{\Gamma}_\alpha$. After solving for the charge current $J_\alpha$, the differential
conductance $G_d$ can be calculated straightly, $G_d=dJ/dV$ with $J_L=-J_R=J$. Accordingly,
the tunneling magnetoresistance ($\mathrm{TMR}$) can be derived by the ratio
$(G_d(0)-G_d(\theta))/G_d(\theta)$ with $G_d(\theta)$
denoting the conductance associated with the angle $\theta$ of the right lead's magnetization.

\section{numerical calculation}\label{numerical}
\begin{figure}[t]
\begin{center}
\includegraphics[width=6.5cm]{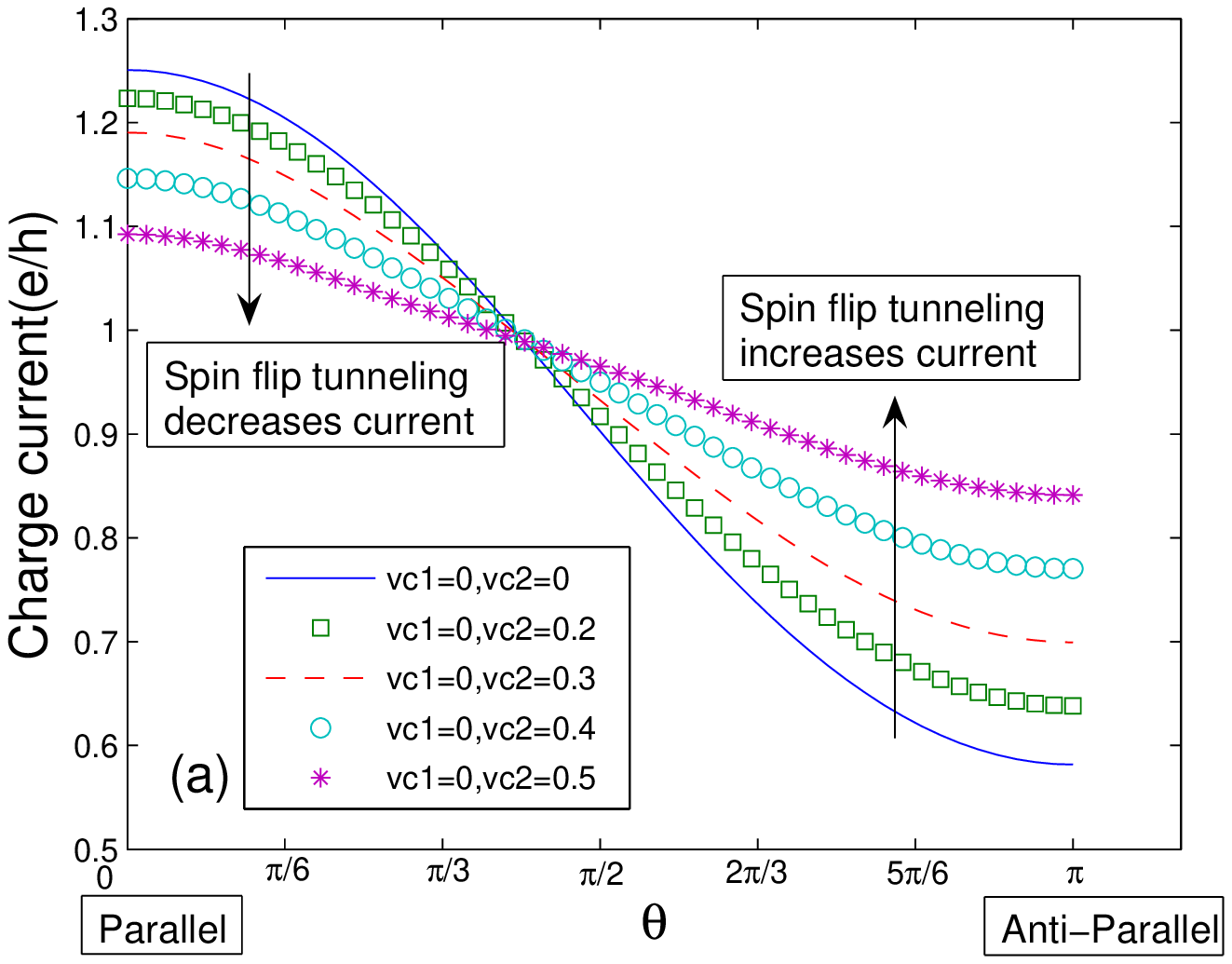}
\includegraphics[width=6.5cm]{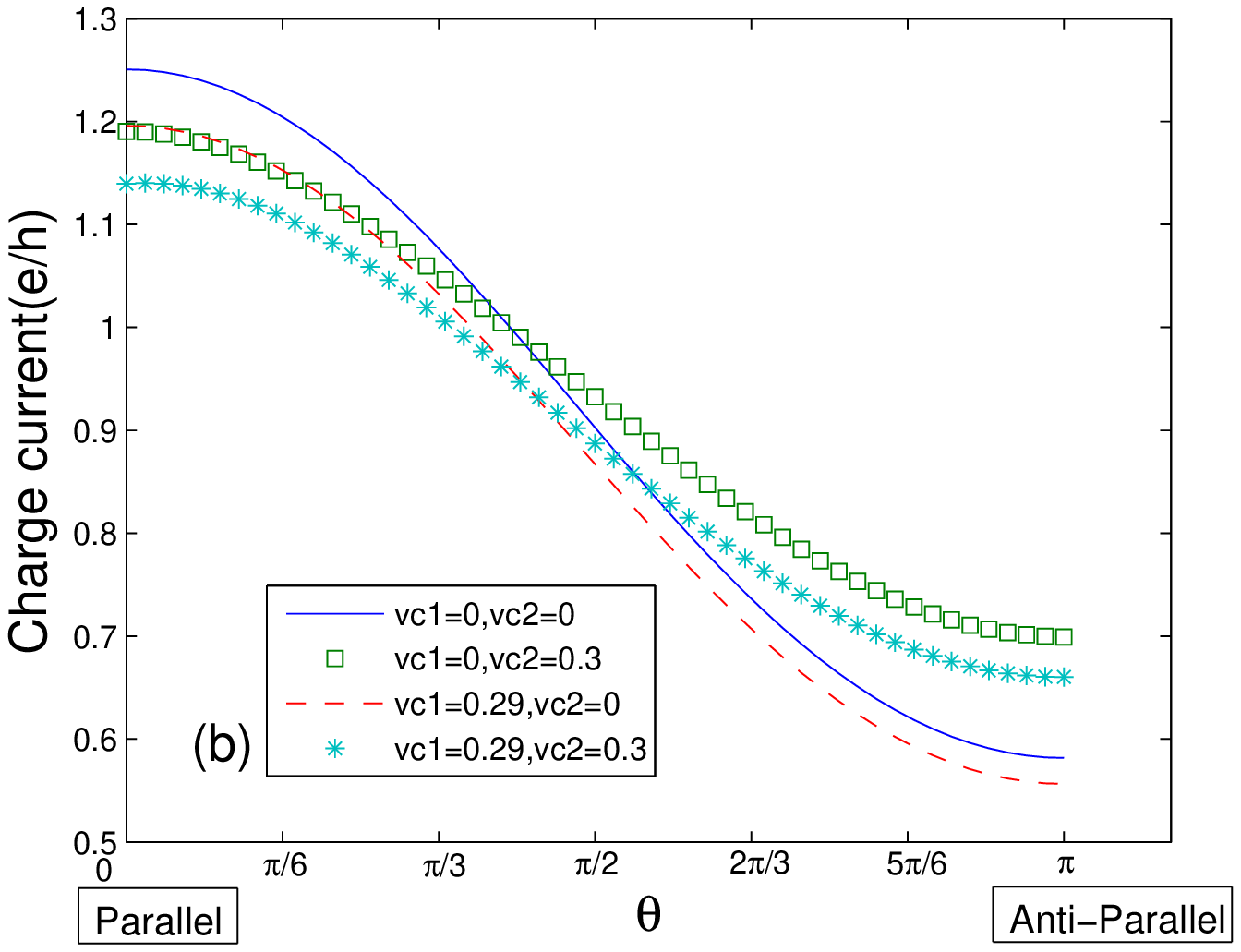}
\caption{\label{fig:current1}   (a) and (b) Charge current $J$ as a function of the angle $\theta$ under different inter-edge
tunnel coupling conditions. In (a), we analyze the role of the spin-flip inter-edge coupling, characterized by $V_{c2}$, while in (b)
we consider the effect of spin-conserving inter-edge coupling, characterized by $V_{c1}$. Other
parameters are $\Gamma_0=1$, $T=0.2$, $p=0.8$, $\mu_0=0.3$ and
$V=0.5$.}
\end{center}
\end{figure}
In this section, we numerically investigate the transport property
of the TI-based device. We adopt the parameters of HgTe quantum wells given in Ref. [\onlinecite{Qi}], i.e., with thickness $d=7\mathrm{nm}$,
$A=364.5 \mathrm{meV ~nm}$, $B=-686 \mathrm{meV ~nm^2}$, $M=-10 \mathrm{meV}$, and $D=512 \mathrm{meV ~nm^2}$, while the edge state velocity
is $v\simeq 5.5\times 10^5$ m/s. The width of the device is set at $W=1000\mathrm{nm}$ so that the energy gap of edge states is very tiny and
can be neglected\cite{Zhou}. The pinching of the two edge states results in a local modification of the spin-orbit coupling, which may induce
spin-conserving and spin-flip tunneling at this narrow region\cite{Zhang,Dolcini,Krueckl}. We adopt the two phenomenological parameters $V_{c1}$
and $V_{c2}$ to describe the two tunneling processes and their relative amplitudes. The Dirac point at equilibrium is set at energy $E=0$, while
the Fermi energy is at $\mu_0=vk_0=0.3\Gamma_0$ ($\Gamma_0=10\mathrm{meV}$). A bias voltage $V$ is applied to the device, such that the chemical potentials
of the leads are $\mu_L=\mu_0+V/2$ and $\mu_R=\mu_0-V/2$.

We first analyze the band dispersion of the edge states in the presence of
the inter-edge tunneling couplings. From the Hamiltonian
$H_C$ of a topological insulator, the energies of top-edge states
are given by $E_\sigma=\pm vk$ in the absence of inter-edge tunneling
couplings. Once the split-gate voltage $V_g$ is applied and
inter-edge tunnel coupling becomes finite, the energies of the edge
states are modified to
\begin{eqnarray}\label{eqn:gap}
E_\sigma'=\pm\sqrt{(vk+V_{c2})^2+V_{c1}^2}.
\end{eqnarray}
The energies of the bottom-edge states can be calculated using the
same method. The spin-preserving part of the inter-edge tunnel
couplings causes a band gap opening of $\Delta=2V_{c1}$. Thus electrons cannot flow through the device
if the Fermi energy is adjusted to lie within the energy range of
$|\mu_0|<V_{c1}$. In this paper, however, we focus on the transport property of
the TI-based device in the regime of $|\mu_0|>V_{c1}$, where there exists
current flow.
\begin{figure}[t]
\begin{center}
\includegraphics[width=6.5cm]{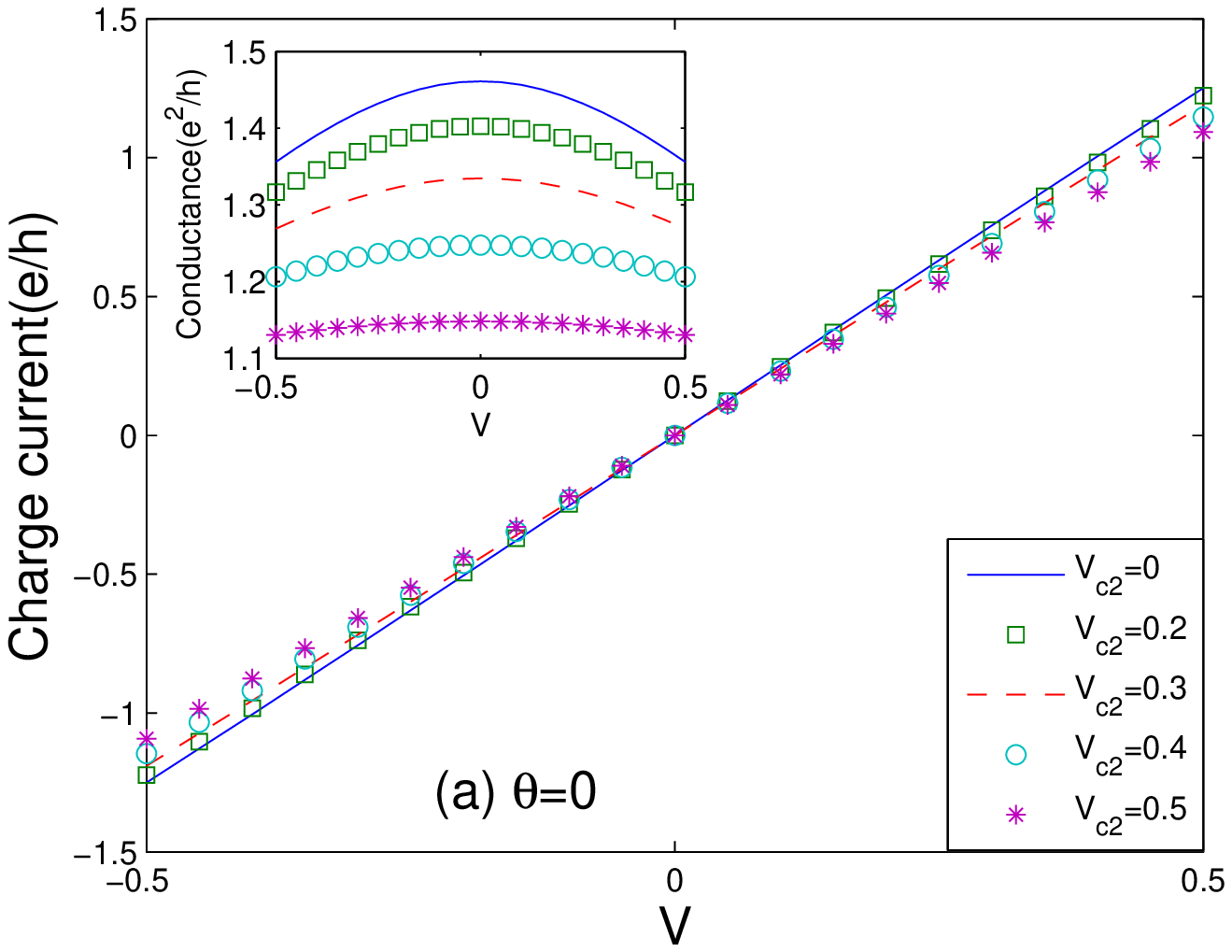}
\includegraphics[width=6.5cm]{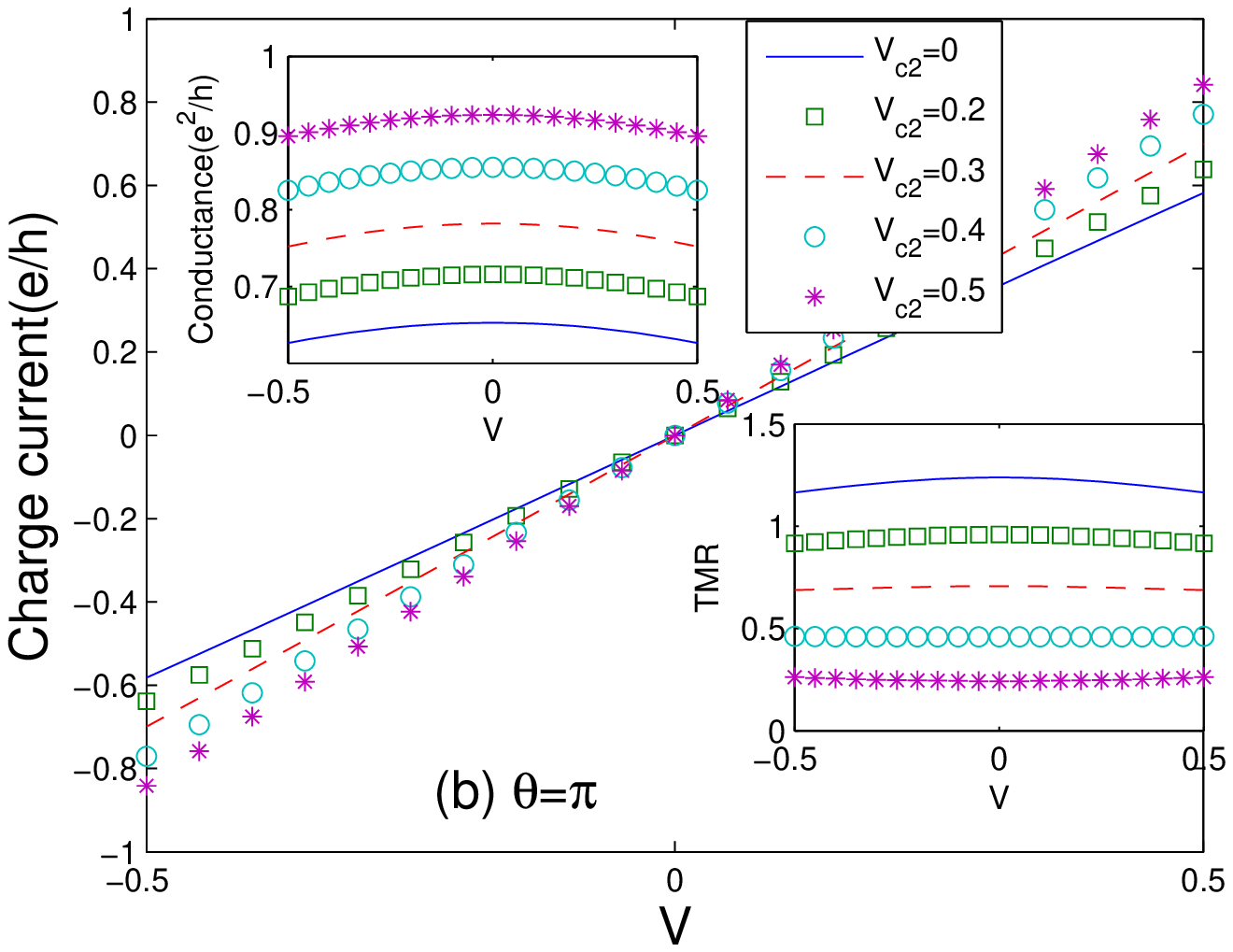}
\caption{\label{fig:current2} Bias dependence of the charge current for the cases (a) $\theta=0$, (b) $\theta=\pi$.
The upper insets in (a) and (b) show the bias dependence of the differential conductance, while the lower inset in (b)
is the corresponding plot of the $\mathrm{TMR}$. $V_{c1}=0$ and the other
parameters are the same as those of Fig.~\ref{fig:current1}.}
\end{center}
\end{figure}
We initially study the effect of the inter-edge tunnel coupling on the
charge current. Fig.~\ref{fig:current1}(a) shows the effect
of the spin-flip tunneling, characterized by the coefficient
$V_{c2}$ on the charge current as a function of magnetization
orientation $\theta$. The charge current decreases with increasing
tunneling strength $V_{c2}$ when $\theta<5\pi/12$, especially at
$\theta=0$ [the parallel configuration]. However, there occurs an opposite
trend, i.e., the charge current can increase with
increasing $V_{c2}$ when $\theta>5\pi/12$. This is especially so when
$\theta=\pi$ [the anti-parallel configuration]. The
observed trend can be explained by considering the spin-state of electrons
in the edge states. In the parallel configuration, spin polarized
electrons can flow through the central TI region along one of the
edge channels in the absence of the inter-edge coupling. For instance,
in the schematic diagram of Fig. 1, we find that when the
magnetization of the leads are pointing in the $z$-direction, then
most of the forward electron flux will flow in the top edge state.
In the presence of inter-edge spin-flip coupling, however, some of
the forward moving electrons in the upper edge will be scattered to
the spin-down state in the lower edge. They will encounter a higher
probability of being blocked at the right electrode since they are a
minority carrier there. On the contrary, for the anti-parallel case,
the current is suppressed in the absence of the inter-edge coupling.
Referring to Figure 1 again, let us assume that the magnetization of
the left (right) leads are magnetized in the $+z$ ($-z$) direction.
The forward spin-up flux in the top edge will be blocked at the
right electrode as spin-up electrons are minority carriers there.
The presence of inter-edge spin-flip coupling (i.e., finite $V_{c2}$)
enables the spin-up electron in the top edge to be scattered into
the forward moving spin-down state in the lower edge. The spin-down
electron will have a higher transmission into the right electrode as
it will be a majority carrier in that electrode. Thus, by utilizing the
presence of spin-flip tunneling at the QPC, one can modulate the transport
properties of this TI-based device for different magnetization
configurations.

At the same time, the spin-conserving inter-edge tunneling coupling at the QPC
induces the back-scattering process. Since this can coexist with
the spin-flip tunneling term, it is deservable to investigate the combined effect of
these two terms on the charge current, as shown in
Fig.~\ref{fig:current1}(b). When there exists only the
spin-conserving tunneling, for example $V_{c1}=0.3$ and $V_{c2}=0$
[red dashed line], the charge current is decreased by the tunneling term via
the back-scattering process. In other words, some of the forward moving electrons in the
upper edge will be scattered to the backward moving spin-up states in the lower edge, thus
decreasing  the overall charge current. The reduction in current due to the back-scattering
is greater for the parallel configuration than that for the antiparallel case.
This is because the current reduction in the parallel case is mainly due to the back-scattering of
majority spin carriers, and this involves more electrons than that of minority electrons in
the anti-parallel case.

In the presence of both types of inter-edge tunnel coupling, i.e. both the spin-flip and
spin-conserving types, the net result will be suppression (enhancement) of charge current
when the device is in the parallel (antiparallel) configuration. In the parallel
configuration, both  the spin-flip and spin-conserving tunneling reduce the current, the
former by scattering to the minority spin state, while the latter by backscattering.
However, in the antiparallel configuration, the two inter-edge couplings have opposite
effects $\--$ the spin-flip (spin-conserving) process increases (decreases) the current.
When the two are of equal strength, e.g., $V_{c1}=V_{c2}=0.3$, the spin-flip process dominates, and so there is a net increase in current in the antiparallel
configuration [see Fig.~\ref{fig:current1}(b)]. The current suppression due to the spin-conserving tunneling also results in the cross-over between net enhancement and
net suppression of charge current occurring at a large-angle configuration beyond $\theta=5\pi/12$
[see Fig.~\ref{fig:current1}(b)]. Thus, the transport properties of our TI-based device can be
effectively modulated by adjusting the relative strengths of the two types of inter-edge tunneling
couplings across the QPC. In order to experimentally verify the above results, one must first prepare
the HgTe-based device with transverse width of at least $W=1000$ nm, as schematically shown in Fig.~\ref{fig:device}.
For the two ferromagnetic leads, the magnetization of the left lead is fixed and pointed in the out-of-plane $\hat{\mathbf{z}}$ direction,
while the right lead is made of a softer magnetic material, so that its magnetization direction can be changed from
$\theta=0$ to $\theta=\pi$ with respect to the left lead, by application of a magnetic field. One can modulate the width of the QPC
by using the gate voltage $V_g$, which would in turn induce different tunneling strengths $V_{c1}$
and $V_{c2}$, due to the finite overlap between edge states on the upper and the lower sides\cite{Krueckl}. By analyzing the voltage
dependence of the charge current, one may determine the variation of the charge current as a function of  $V_{c1}$ and $V_{c2}$.

Next, we analyze the charge current $J$ and the conductance $G_d$ of the TI-based device under varying bias voltage. In
Fig.~\ref{fig:current2}(a) and (b), the bias dependence of $J$ is plotted for different spin-flip coupling
strengths $V_{c2}$, for the parallel $(\theta=0)$, and antiparallel $(\theta=\pi)$ configurations.
As noted earlier, for the parallel (antiparallel) configuration, \textbf{$J$} decreases (increases) as $V_{c2}$
increases. Similarly, the bias dependence plot of $G_d$ also shows a decrease (increase) with increasing $V_{c2}$ when $\theta=0$ ($\theta=\pi$),
as shown in the upper insets of Fig. 3(a) [(b)]. This may be explained by the fact that a finite $V_{c2}$ would lead to coupling of two edge
states with opposite spins, resulting in spin mixing and thus reducing the degree of spin asymmetry in the device. Consequently, the TMR ratio
decreases with increasing $V_{c2}$, as shown by the lower inset of Fig.~\ref{fig:current2}(b). For both configurations, the conductance
also decreases with increasing bias voltage $V$, with a larger decrease occurring for the parallel case.
This translates to a slight decrease in TMR ratio with bias voltage $V$ for small coupling strength $V_{c2}$.
Hence, the conductance and the TMR of the TI-based device can be modulated by both the source-drain voltage $V$, and
the split-gate voltage $V_g$, which alters the coupling strength $V_{c2}$.\\

\section{Conclusion}\label{summary}
In summary, we have investigated the transport of a two-dimensional topological insulator (TI) sandwiched
between two ferromagnetic (FM) leads, and incorporating a quantum point contact (QPC) which couples the edge states along
the top and bottom boundaries. The conductance across the device is calculated via the nonequilibrium Green's function (NEGF)
method. We found that spin-flip tunnel coupling between the edge states mediated by the QPC can enhance (suppress) the
current when the magnetization of the two FM leads is in the antiparallel (parallel) configuration. On the other hand,
the spin-conserving tunnel coupling leads to backscattering which reduces the overall current regardless of the
magnetization alignment of the FM leads. The conductance and TMR of the TI-based device can be controlled by both the
source-drain voltage $V$, as well as the split-gate voltage $V_g$, the latter modifying the inter-edge coupling strengths
across the QPC. Our proposed device suggests a new class of TI-based devices which incorporate a QPC to
electrically control the inter-edge coupling, so as to harness the topological and spin-polarized nature of the TI edge
states for practical application.

\acknowledgments We greatfully acknowledge the National
University of Singapore (NUS) Grant No. R-263-000-632-592 and the SMF-NUS Research Horizons Award for financially
supporting their work. The work was also supported by Innovation
Research Team for Spintronic Materials and Devices of Zhejiang
Province.


\begin{thebibliography}{99}
\bibitem{Kane}
C. L. Kane and E. J. Mele, Phys. Rev. Lett. {\bf95}, 146802 (2005).
\bibitem{Bernevig}
B. A. Bernevig and S. C. Zhang, Phys. Rev. Lett. {\bf96}, 106802
(2006).
\bibitem{Hsieh}
D. Hsieh, D. Qian, L. Wray, Y. Xia, Y. S. Hor, R. J. Cava, and M. Z.
Hasan, Nature {\bf452}, 970 (2008).
\bibitem{Zhang}
H. J. Zhang, C. Liu, X. Qi, X. Dai, Z. Fang, and S. Zhang, Nature Phys. {\bf5},
438 (2009).
\bibitem{Xia}
Y. Xia, D. Qian, D. Hsieh, L. Wray, A. Pal, H. Lin, A. Bansil, D. Grauer,
Y. S. Hor, and R. J. Cava, Nature Phys. {\bf5}, 398 (2009).
\bibitem{Qi}
X. L. Qi and S. C. Zhang, Rev. Mod. Phys. {\bf83}, 1057 (2011).
\bibitem{Hossain}
M. Z. Hossain, S. L. Rumyantsev, K. M. F. Shahil, D. Teweldebrhan, M. Shur,
and A. A. Balandin, ACS Nano 5, 2657 (2011).
\bibitem{Shahil}
K. M. F. Shahil, M. Z. Hossain, D. Teweldebrhan, and A. A. Balandina,
Appl. Phys. Lett. {\bf96}, 153103 (2010).
\bibitem{Fujita}
T. Fujita, M. B. A. Jalil, and S. G. Tan, Appl. Phys. Express {\bf4}, 094201 (2011).
\bibitem{Moore}
J. E. Moore, Nature 464, 194 (2010).
\bibitem{Tang}
H. Tang, D. Liang, R. L. J. Qiu, and X. P. A. Gao, ACS Nano {\bf5}, 7510 (2011).
\bibitem{Teweldebrhan}
D. Teweldebrhan, V. Goyal, and A. A. Balandin, Nano Lett. {\bf10}, 1209 (2010).
\bibitem{Teweldebrhan2}
D. Teweldebrhan, V. Goyal, M. Rahman, and A. A. Balandin, Appl. Phys. Lett. {\bf96}, 053107 (2010).
\bibitem{Goyal}
V. Goyal, D. Teweldebrhan, and A. A. Balandina, Appl. Phys. Lett. {\bf97}, 133117 (2010).
\bibitem{Bernevig2}
B. A. Bernevig, T. L. Hughes, and S. C. Zhang, Science {\bf314},
1757 (2006).
\bibitem{Konig2}
M. Konig, S. Wiedmann, C. Brune, A. Roth, H. Buhmann, L. W.
Molenkamp, X. L. Qi, and S. C. Zhang, Science {\bf318}, 766 (2007).
\bibitem{Roth}
A. Roth, C. Br\"{u}ne, H. Buhmann, L. W. Molenkamp, J. Maciejko, X.
L. Qi, and S. C. Zhang, Science {\bf330}, 1746 (2009).
\bibitem{Buttiker}
M. Buttiker, Science {\bf325}, 278 (2009).
\bibitem{Fu}
L. Fu and C. L. Kane, Phys. Rev. B {\bf76}, 045302 (2007).
\bibitem{Murakami}
S. Murakami, Phys. Rev. Lett. {\bf97}, 236805 (2006).
\bibitem{Liu}
C. X. Liu, T. L. Hughes, X. L. Qi, K. Wang, and S. C. Zhang, Phys.
Rev. Lett. {\bf100}, 236601 (2008).
\bibitem{Knez}
I. Knez, R. R. Du, and G. Sullivan, Phys. Rev. B {\bf81}, 201301(R) (2010).
\bibitem{Shitade}
A. Shitade, H. Katsura, J. Kune\v{s}, X. L. Qi, S. C. Zhang, and N.
Nagaosa, Phys. Rev. Lett. {\bf102}, 256403 (2009).
\bibitem{Wada}
M. Wada, S. Murakami, F. Freimuth and G. Bihlmayer, Phys. Rev. B {\bf83}, 121310(R)
(2011).
\bibitem{Zhou}
B. Zhou, H. Z. Lu, R. L. Chu, S. Q. Shen, and Q. Niu, Phys. Rev.
Lett. 101, 246807 (2008).
\bibitem{Hou}
C.Y. Hou, E.A. Kim, and C. Chamon, Phys. Rev. Lett. {\bf102},
076602 (2009).
\bibitem{Strom}
A. Strom, and H. Johannesson, Phys. Rev. Lett. {\bf102}, 096806 (2009).
\bibitem{Tanaka}
Y. Tanaka, and N. Nagaosa, Phys. Rev. Lett. {\bf103}, 166403 (2009).
\bibitem{Teo}
J. C. Y. Teo, and C. L. Kane, Phys. Rev. B {\bf79}, 235321 (2009).
\bibitem{Zyuzin}
V. A. Zyuzin, and G. A. Fiete, Phys. Rev. B {\bf82}, 113305 (2010).
\bibitem{Zhang2}
L. B. Zhang, F. Cheng, F. Zhai, and K. Chang, Phys. Rev. B {\bf83},
081402 (2011).
\bibitem{Dolcini}
F. Dolcini, Phys. Rev. B {\bf83}, 165304 (2011).
\bibitem{Krueckl}
V. Krueckl and K. Richter, Phys. Rev. Lett. 107, 086803 (2011).
\bibitem{Maciejko}
J. Maciejko, E. A. Kim, and X. L. Qi, Phys. Rev. B {\bf82}, 195409
(2010).
\bibitem{Meir}
Y. Meir and N. S. Wingreen, Phys. Rev. Lett. {\bf68}, 2512 (1992);
A. P. Jauho, N. S. Wingreen, and Y. Meir, Phys. Rev. B {\bf50}, 5528
(1994).
\end{thebibliography}
\end{document}